\documentclass[10pt,twocolumn]{IEEEtran}
\usepackage{xspace,amsthm,amsmath,amssymb,amsfonts,epsfig,syntonly}
\usepackage{cite,bm,color,url,textcomp,balance}
\usepackage{epstopdf}
\usepackage{empheq}

\usepackage{algorithm}
\usepackage{algorithmicx}
\usepackage{algpseudocode}

\usepackage{graphicx}{\large}
\usepackage{subfigure}

\usepackage{amsmath,amssymb,amsthm}
\usepackage{bigints}
\usepackage{stfloats,siunitx}

\newcommand*{\blue}{\color{black}}

\long\def\symbolfootnote[#1]#2{\begingroup
\def\thefootnote{\fnsymbol{footnote}}
\footnote[#1]{#2}\endgroup}

\psfull

\allowdisplaybreaks[4]

\IEEEoverridecommandlockouts

\begin{document}
\title{Trajectory Planning of Cellular-Connected UAV for Communication-assisted Radar Sensing}

\author{Shuyan Hu,~\IEEEmembership{Member, IEEE}, Xin Yuan,~\IEEEmembership{Member, IEEE},
Wei Ni,~\IEEEmembership{Senior Member, IEEE}, \\ and Xin Wang,~\IEEEmembership{Senior Member, IEEE}
\thanks{
Copyright (c) 2015 IEEE. Personal use of this material is permitted. However, permission to use this material for any other purposes must be obtained from the IEEE by sending a request to pubs-permissions@ieee.org.
Shuyan Hu and Xin Yuan contributed equally to this work.
(Corresponding author: Xin Wang).

S. Hu is with the Key Lab of EMW Information (MoE), the School of Information Science and Technology, Fudan University, Shanghai 200433, China (e-mail: syhu14@fudan.edu.cn).

X. Yuan and W. Ni are with the Data61, Commonwealth Scientific and Industrial Research Organization, Sydney, NSW 2122, Australia
(e-mails: \{xin.yuan, wei.ni\}@data61.csiro.au).

X. Wang is with the Key Lab of EMW Information (MoE), the Department of Communication Science and Engineering, Fudan University, Shanghai 200433, China (e-mail: xwang11@fudan.edu.cn).
}}

\markboth{ACCEPTED BY IEEE TRANSACTIONS ON COMMUNICATIONS}%
{HU \MakeLowercase{\textit{et al.}}: Trajectory Planning of Cellular-Connected UAV for Communication-assisted Radar Sensing}



\maketitle

\begin{abstract}
Being a key technology for beyond fifth-generation wireless systems, joint  communication and radar sensing (JCAS) utilizes
the reflections of communication signals to detect foreign objects and deliver situational awareness.
A cellular-connected unmanned aerial vehicle (UAV) is uniquely suited to form a mobile bistatic synthetic aperture radar (SAR)
with its serving base station (BS) to sense over large areas with superb sensing resolutions
at no additional requirement of spectrum.
This paper designs this novel BS-UAV bistatic SAR platform, and optimizes the flight path of the UAV to minimize its propulsion energy
and guarantee the required sensing resolutions on a series of interesting landmarks.
A new trajectory planning algorithm is developed to convexify the propulsion energy and resolution requirements by using successive convex approximation and block coordinate descent.
Effective trajectories are obtained with a polynomial complexity.
Extensive simulations reveal that the proposed trajectory planning algorithm outperforms significantly its alternative that minimizes
the flight distance of cellular-aided sensing missions in terms of energy efficiency and effective consumption fluctuation.
The energy saving offered by the proposed algorithm can be as significant as 55\%.

\end{abstract}


\section{Introduction}
Recent advancements in the fifth-generation and beyond (5G/B5G) networks are increasingly enabling extensive background situation-
and position-aware smart applications, including autonomous driving, distant medical care, and smart industry~\cite{hu21dml}.
5G/B5G networks are also envisioned to offer high-resolution sensing in support of these applications~\cite{kumari18}.
For this reason, joint communication and radar sensing (JCAS) has been deemed as one of the essential technologies in
5G/B5G systems~\cite{zhang21}.
By integrating radio communication and sensing into a single system, JCAS measures the reflections of communication signals
to sense the location, velocity, and feature signal of targets and motions~\cite{sturm11, luo19}.
{\blue
This is different from the widely accessible Global Positioning System (GPS), which enables every GPS receiver to locate itself by correlating the preambles/pilots from multiple (four or more) satellites~\cite{stotts, Humphreys}. Neither of the GPS satellites and ground receivers sense the environment or deliver situational awareness, as opposed to JCAS~\cite{zhang21}.}

Cellular-connected unmanned aerial vehicles (UAVs) have been increasingly considered for their operability and applicability to UAV operations over wide areas~\cite{zeng21xu, yuan20}.
Apart from being an aerial user of communication services, a cellular-connected UAV can potentially utilize the cellular signals
from corresponding base stations (BSs),
and collect the echo signals from the ground (e.g., metallic) objects to perform radar sensing in areas covered by the BSs.
Given the excellent mobility, the UAV can potentially form a bistatic synthetic aperture radar (SAR) with its serving BS to sense the environment.
While dedicated sensing or radar techniques may detect objects with even better resolution (e.g., because of their wider signal bandwidth and higher signal power), they inevitably bring radio radiation and pollution~\cite{zhang21}.
Nevertheless, the UAVs are passive receivers, and neither increase the radio footprint of the system nor produce radio pollution.

{\blue
}

Fig.~\ref{system} illustrates the new cellular-assisted radar sensing by a cellular-connected UAV,
referred to as BS-UAV bistatic SAR sensing, 
where the reflections of the downlink cellular signals are picked up by the UAV and used to produce radio/radar images for object detection.
Since the UAV is typically power-constrained, it is critical to carefully plan its trajectory to deliver energy-efficient radar sensing missions and extend mission time.
Existing works have only designed the trajectory to minimize the UAV propulsion energy for wireless communication services,
such as~\cite{zyong, sun21, hu20, hu21}.
None has considered the energy consumption in the new scenario of cellular-assisted radar sensing with a cellular-connected UAV.

\subsection{Related Work}\label{sec: related work}

Existing studies of JCAS systems have typically focused on the beamforming design at the BSs,
and have not considered the use of a UAV.
These JCAS systems are in essence monostatic radars, which integrate the transmitter and receiver on the same platform.
In~\cite{liu18twc}, waveform design was optimized to confirm the produced signal
to the expected sensing waveform subject to the signal-to-interference-and-noise ratio (SINR) requirement for multiuser
multiple-input multiple-output (MIMO) communications.
In~\cite{liu18tsp}, globally optimal waveforms were devised for several expected sensing beam patterns to minimize multiuser interference.
In~\cite{zhang19tvt}, a multibeam technique was introduced for millimeter-wave (mmWave) JCAS with analog antenna arrays,
and a fixed transmission subbeam was produced together with direction-changing scanning subbeams across various packets.
In~\cite{zhitong21}, a multi-metric waveform was designed for multiple-input single-output (MISO) communications
to maximize the SINR at the users.
On the other hand, a higher signal power is typically on demand for high-resolution sensing over long ranges,
whereas the BSs usually have a limited transmit power.
The sensing performance may considerably degrade if the targets are far away from the BS,
because of the significant round-trip path loss attenuation of the echoed signals.

UAVs have been emerging as a new type of passive receiver in bistatic radar systems.
Unlike a monostatic radar, a bistatic radar uses geographically separated antennas for signal transmission and reception.
The German TanDEM-X project (initiated in June 2010) developed the first prototype of a space-borne bistatic radar, 
relying on twin low-Earth-orbiting (LEO) satellites traveling in near formation~\cite{tandem10}.
The High-Resolution Wide-Swath (HRWS) project was proposed to be the next-generation German space-borne radar platform for
geological and geographical observation beyond 2030, by leveraging the formation movement of an active satellite
and three passive satellites~\cite{bartusch21}.
Mounted on mobile platforms,
a bistatic SAR can perform HRWS sensing by making use of the relative movements between the SAR and the targets.
The small-sized real antenna aperture can realize the capability of a larger aperture radar through data focusing
and signal processing~\cite{wang10}.
A clutter suppression and mobile target imaging strategy was developed in~\cite{zhang21jst} for a GEO-LEO bistatic
SAR system, which was robust against the quickly traveling object with the Doppler centroid ambiguity.

With a geosynchronous (GEO) SAR satellite as the transmitter, a UAV can receive the echo signals from terrestrial targets
and achieve two-dimensional (2D) imaging.
The flight path of the UAV receiver was planned in~\cite{sun16} for such a GEO-UAV bistatic SAR system to sense rough terrains.
A range model was developed in~\cite{sun21sar} to address the distinctiveness
of the GEO-UAV bistatic SAR echo and the variations of the orbit positions of the GEO transmitter.
A motion compensation algorithm was proposed in~\cite{wang22} for a mini-UAV-based bistatic SAR system
to address the perturbation and spatial variance of the platform.
However, none of the existing works [14]--[24] have considered the BS-UAV bistatic SAR, as considered in this paper.
None of the existing studies have considered the energy consumption of a UAV-based bistatic SAR system.
{\blue The works in [22]--[24] were focused on data focusing or image processing algorithms, where the geosynchronous satellites 
transmitted dedicated signals for ground object detection and did not transmit any data. 
The energy consumption of the UAV was not considered in [22] for trajectory planning,
and the UAV trajectory design was not addressed in [23] and [24].
The results of [22]--[24] cannot apply directly to a BS-UAV bistatic SAR system for energy-efficient trajectory design. }

\subsection{Contribution and Organization}
In this paper, we propose a novel framework for cellular-aided radar sensing with a cellular-connected UAV,
i.e., BS-UAV bistatic SAR sensing,
where the wireless transmissions of a BS serve as the excitation signals of a bistatic SAR system,
and the UAV collects the echoed signals for radar sensing.
Given a series of landmarks to be sensed/observed, the UAV's trajectory is optimized to minimize its propulsion energy
while satisfying the range and azimuth resolutions of sensing,
thereby delivering energy-efficient BS-UAV bistatic SAR sensing missions and extending mission durations.
This trajectory planning problem is non-convex because of the non-convexity of its objective function and constraints.
We convexify the problem and obtain a quality solution efficiently by utilizing successive convex approximation (SCA)
and block coordinate descent (BCD).

The key contributions of the paper are listed as follows.
\begin{itemize}
\item A novel BS-UAV bistatic SAR platform is proposed to reuse cellular downlink signals as excitation signals.
A cellular-connected UAV receives the reflected signals and conducts radar sensing on landmarks.

\item The range and azimuth resolutions of the BS-UAV bistatic SAR are analyzed.
An effective sensing area of the BS-UAV bistatic SAR is established to effectively capture a landmark.

\item The UAV's trajectory is optimized to minimize its propulsion energy and ensure that all landmarks are captured
in the effective sensing areas with the required range and azimuth sensing resolutions. 
By applying the SCA and BCD methods, a new algorithm is developed to deliver energy-efficient trajectories
with a polynomial complexity.
\end{itemize}
Extensive simulations corroborate that, in terms of energy efficiency and energy usage fluctuations,
the proposed trajectory planning algorithm outperforms a baseline scheme that minimizes the flight distance of the UAV
on the cellular-aided radar sensing mission.
The proposed algorithm can save up to $55\%$ of the UAV's propulsion energy, as compared to the baseline scheme.

The remainder of this paper is arranged as follows.
Section~\ref{sec.model} describes the system model.
Section~\ref{sec.prob} formulates the UAV trajectory design problem for the BS-UAV bistatic SAR,
and elaborates on the proposed trajectory planning algorithm.
Performance evaluations are conducted in Section~\ref{sec.sim}, followed by concluding remarks in Section~\ref{sec.con}.

\begin{figure*}
\centering
\includegraphics[width=0.8\textwidth]{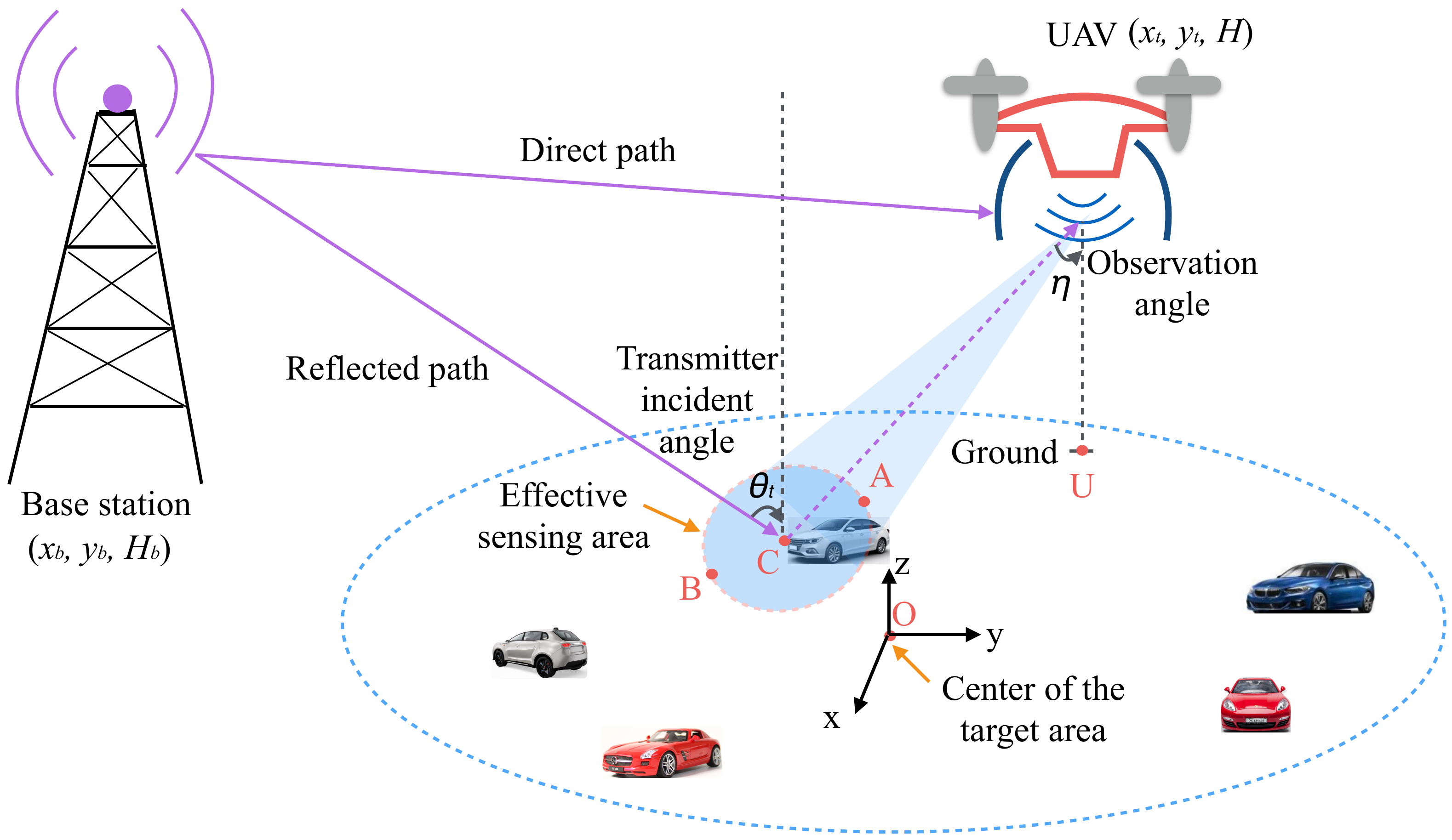}
\caption{System model of joint communication and radar sensing by a BS-UAV bistatic SAR, where the UAV receives the downlink cellular signals from the BS and the reflections of the signals reflected by (metalic) objects.
The reflection signals can be extracted and used to produce radio/radar images and detect the objects.}
\label{system}
\end{figure*}

\section{System Model}\label{sec.model}
In the considered system, a BS is deployed to deliver downlink wireless services to cellular users.
A cellular-connected UAV equipped with a side-looking SAR (receiver) flies and collects the echoes (or reflections) of the communication
signals originated from the BS, to sense (metallic) objects and gain situational awareness; see Fig.~\ref{system}.
The signals are buffered at the UAV, and offloaded and post-processed (at the BS) after the sensing mission.
Given the knowledge of the communication signals, the direct path from the BS to the UAV can be canceled.
The remaining echo signals can be used to sense and decide the number and locations of the objects.
In this way, the UAV can perform the sensing task by utilizing the communication signals,
without the need of dedicated sensing signals, and thereby reduces the radio fingerprint and carbon emission.

{\blue 
The proposed UAV-based sensing platform is insusceptible to inter-cell interference, since it does not need to detect or decode data signals. The transmissions from other BSs can serve as additional illuminating sources to enhance the reflection on the objects. 
The UAV may utilize the pilot signals of the BSs to extract the reflections of objects~\cite{ghosh}. 
With the knowledge of the pilot signals, the collected reflection signals by the UAV can be correlated to cancel the direct paths from any BSs and extract the reflections by the objects to detect the objects. 
Alternatively, the UAV may store the captured signals, and only start to process the signals after it returns from the mission and has access to the transmitted signals of the BSs. Then, correlation can be conducted between the captured signals and the transmitted signals to extract the reflections and detect the objects.
On the other hand, the proposed trajectory planning algorithm can be performed at the UAV with limited assistance of the BSs. It only requires the information of the landmarks prior to the trajectory planning. The inter-cell interference has little impact on the algorithm, when the algorithm is in operation. }

\subsection{Object Distribution}
Consider a 2D ground region $\mathcal A$ with radius $R$.
We assume that $\mathcal A$ is inside the coverage area of the BS.
The number of illuminating (e.g., metallic) objects in $\mathcal A$, denoted by $N(\mathcal A)$,
follows the homogeneous spatial Poisson point process (SPPP).
{\blue The reason is that the points are uniformly distributed in the circular area (with the same density $\lambda$).}
The probability density function (PDF) is given by~\cite{stochgeo12}
\begin{equation}
\Pr (N(\mathcal A)=n) = \frac{\lambda | \mathcal A |^n}{n!} e^{-\lambda | \mathcal A |}, n=0,1, \ldots,
\end{equation}
where $\lambda >0$ is the intensity, and $| \mathcal A |$ is the Lebesgue measure for the size of the region.
The objects are uniformly randomly distributed in $\mathcal A$.

Let $d_i$, $i=1,\cdots,N(\mathcal A)-1$, collect the Euclidean distances from an arbitrarily selected (and designated) object to
the other $(N(\mathcal A)-1)$ objects within the disk region with radius $R$; 
and assume that $d_i, \,i=1,\cdots,N(\mathcal A)-1$, are independent and identically distributed (i.i.d.), non-negative random variables.
{\blue Given the uniform stationary distribution of the objects in the disk region with radius $R$,
the cumulative density function (CDF) of $d_i,\,i=1,\cdots,N(\mathcal A)-1$ is given by~\cite{2018Yuan}}\footnote{{\blue
Given the uniform stationary distribution of the objects in the disk region with radius $R$, the PDF of $d_i,\,i=1,\cdots,N(\mathcal A)-1$ is given by~\cite{2018Yuan}
\begin{equation}
f_{d_i}(l)\!=\!\frac{2l}{R^2}\left(\frac{2}{\pi}\cos^{-1}\!\left(\!\frac{l}{2R}\right)\!\!-\!\frac{l}{\pi R }\sqrt{1\!-\!\frac{l^2}{4R^2}}\right),
0 \leq l \leq 2R. \notag
\end{equation} }}
\begin{equation}\label{eq:cpdf_of_l}
\begin{aligned}
F_{d_i}(l) & = 1 + \frac{2}{\pi} \left(\frac{l^2}{R^2} - 1 \right) \cos^{-1}\left(\frac{l}{2R}\right) \\
         & \quad - \frac{1}{\pi R}\left(1 + \frac{l^2}{2R^2} \right)\sqrt{1-\frac{l^2}{4R^2}},\;0 \leq l \leq 2R.
\end{aligned}
\end{equation}

Let $d_{\min}$ denote the shortest of the distance between two objects, i.e., $d_{\min}=\min_{i=1,\cdots,N-1}\{d_i\}$.
By exploiting Order Statistics, the CDF of $d_{\min}$ can be given by
\begin{equation}\label{cdf_of_lmin}
\begin{aligned}
F_{d_{\min}}(l)&=1-\left[1-F_{d_i}(l)\right]^{N(\mathcal A)-1}\\
&=1-\left[\frac{1}{\pi R}\left(1 + \frac{l^2}{2R^2} \right)\sqrt{1-\frac{l^2}{4R^2}} \right.\\
& \left.\qquad\quad - \frac{2}{\pi} \left(\frac{l^2}{R^2} - 1 \right) \cos^{-1}\left(\frac{l}{2R}\right) \right]^{N(\mathcal A)-1}.
\end{aligned}
\end{equation}
{\blue The resolution of the proposed UAV-based bistatic SAR platform is set to be finer than the expected shortest distance between two neighboring objects, i.e., $d_{\min}$, to reasonably distinguish any two different objects. 
For this reason, we derive the expectation of $d_{\min}$ 
based on the CDF in \eqref{cdf_of_lmin}.
The expectation is an integral over $l$, the distance between any two objects in the circular coverage area of the BS with the radius of $R$. Here, $0 \le l \le 2R$.
Then, the}
expectation of ${d_{\min}}$ can be given by
\begin{equation}\label{expectation_of_lmin}
\begin{aligned}
& \mathbb{E}\left[{d_{\min}}\right]  = \int_{0}^{2R} l \; d{F_{d_{\min}}(l)}\\
& =- l\Big[\!-F_{d_i}(l)\Big]^{N(\mathcal A)-1} \bigg|_{0}^{2R}
 +\int_{0}^{2R}\Big[1-F_{d_i}(l)\Big]^{N(\mathcal A)-1}dl\\
& = \int_{0}^{2R}\left[\frac{1}{\pi R}\left(1 + \frac{l^2}{2R^2} \right)\sqrt{1-\frac{l^2}{4R^2}} \right.\\
& \left.\qquad\quad - \frac{2}{\pi} \left(\frac{l^2}{R^2} - 1 \right) \cos^{-1}\left(\frac{l}{2R}\right) \right]^{N(\mathcal A)-1}dl,
\end{aligned}
\end{equation}
where $F_{d_i}(0) = 1-\frac{1}{\pi R}$ and $F_{d_i}(2R) = 1$.

\subsection{UAV Mobility}
{\blue In practice, the receiver of a bistatic SAR system flies at a constant altitude, as a stable flight can stabilize the accuracy of the SAR~\cite{zhangyun}. For this reason, we consider a horizontally flying UAV at a constant altitude, $H$ (in meters), in line with practical implementations, and focus this paper on 2D trajectory planning.
Nevertheless, the algorithm developed in this paper can be potentially extended to three-dimensional (3D) trajectory planning with varying UAV altitudes $H_t, \forall t$, as will be discussed in Section III.}

The UAV executes the sensing mission for a fixed time horizon of $T$ seconds, which is divided
into $T_w$ time slots indexed by $t$, and $t=1, \ldots, T_w$.
The duration of a time slot $t$ is $\delta$~seconds.
The UAV's time-varying 2D coordinates are $\mathbf q_t=[x_t,y_t]^T, \forall t$,
with the initial location $\mathbf q_0=[x_0, y_0]^T$.
Let $V_t$ denote the UAV speed at time slot $t$, which is upper bounded by $V_{m}$.
The UAV's mobility constraint is given by
\begin{equation}\label{eq.mob}
\|\mathbf q_t- \mathbf q_{t-1} \| = \delta V_{t}  \leq \delta V_{m}, ~\forall t.
\end{equation}

For a rotary-wing UAV at a speed $V_t$,
the propulsion power at time slot $t$, denoted by $P_t$, is shown by~\cite{zyong}
\begin{equation}\label{speed}
\begin{aligned}
P_t = & P_0 \left(1+\frac{3{V_t}^2}{U_{tip}^2}\right) + P_1\left(\sqrt{1+\frac{{V_t}^4}{4v_0^4}}
-\frac{{V_t}^2}{2 v_0^2}\right)^{\frac{1}{2}}\\
&+ \frac{1}{2} d_f\rho s A{V_t}^3,
\end{aligned}
\end{equation}
where $P_0$ and $P_1$ stand for the fixed {\it blade} and {\it induced powers} when the aircraft hovers,
respectively; $U_{tip}$ denotes the rotor velocity; $v_0$ is the mean rotor velocity when the aircraft hovers;
$d_f$ and $s$ stand for the fuselage drag fraction and rotor solidity, respectively;
and $\rho$ and $A$ are the gaseous density and rotor disc area, respectively.

\subsection{Airborne Bistatic Side-Looking SAR}

As shown in Fig.~\ref{system}, the effective sampling SAR sensing area of the UAV, denoted by $\mathcal U_t$,
is an ellipsoid centered at point $C$ as per slot $t$.
We assume that the effective sensing area $\mathcal U_t$ is always to the right of the UAV,
by considering a side-looking SAR~\cite{marc10}.
The major axis of the ellipsoid $\mathcal U_t$, $L_{\rm AB}$, is perpendicular to the UAV's heading.
Let $\alpha_t \in [0, 2 \pi)$ denote the angle between the UAV's heading and the $x$-axis at time slot $t$.
We have
\begin{equation}\label{eq.sincos}
\sin \alpha_t =  \frac{y_t - y_{t-1}}{\delta V_t};~\cos \alpha_t = \frac{x_t - x_{t-1}}{\delta V_t}.
\end{equation}
Further let $\eta \in (0, \pi /2)$ denote the observation angle of the side-looking SAR at the UAV,
which is fixed over time.
Given the horizontal coordinates of the UAV, i.e., $U=(x_t, y_t)$ at slot $t$, and $L_{\rm UC} = H \tan \eta$,
the coordinates of the center $C$, i.e., $\mathbf q_c^t=[x_c^t, y_c^t]^T$, can be written as
\begin{equation}
x_c^t = x_t + H \tan \eta \sin \alpha_t; ~y_c^t = y_t - H \tan \eta \cos \alpha_t.
\end{equation}
To guarantee the sensing performance, $C$ should always be inside the target area $\mathcal A$.
We have $L_{\rm OC} \le R,~\forall t$.

At any time slot $t$, the range resolution $\delta_r^t$ and the azimuth resolution $\delta_a^t$ of the BS-UAV bistatic SAR are given
by~\cite[Eq. (5)]{marc10}, \cite[Eqs. (24) \& (29)]{moccia11}
\begin{subequations}\label{eq.resolution}
\begin{align}
\delta_r^t &= \frac{c}{B (\sin \eta + \sin \theta_t)}; \\
\delta_a^t &= \frac{\lambda_c H}{T_d V_t \cos \eta}, \label{eq.azimres}
\end{align}
\end{subequations}
where $c$ is the speed of light, $\lambda_c$ is the wavelength, $B$ is the transmit signal bandwidth of the BS,
$\theta_t \in (0, \pi /2)$ is the incidence angle of the transmitted signal (from the BS) with respect to the point $C$,
and $T_d$ is the coherent SAR integration time~\cite{ulaby82}.
Given the horizontal coordinates of the BS $\mathbf q_b=[x_b, y_b]^T$ and its height $H_b$, we have
\begin{equation}
\sin \theta_t = \frac{\|\mathbf q_c^t- \mathbf q_b \|}{\sqrt{ \|\mathbf q_c^t- \mathbf q_b \| ^2+ {H_b}^2}}.
\end{equation}

In order to distinguish the objects, the resolution of the SAR has to be finer than the shortest expected distance between any two objects,
$d_{\min}$, i.e., $\delta_r^t \le d_{\min}$ and $\delta_a^t \le d_{\min}$.
{\blue This resolution requirement is used to detect and differentiate different objects by leveraging the typically narrow signal bandwidths of cellular systems and, in turn, the poor resolution of sensing~\cite{cudak}. 
It is worth mentioning that 
we do not consider object identification in this paper, as it would require wider signal bandwidths to achieve better resolutions for detecting the shape of an object or even imaging the object~\cite{moccia11}.
Nevertheless, the proposed trajectory planning algorithm would still be relevant under a wider signal bandwidth and correspondingly a finer resolution requirement.
The algorithm is generic and suitable for different resolution requirements. }

\begin{figure}
\centering
\includegraphics[width=0.5\textwidth]{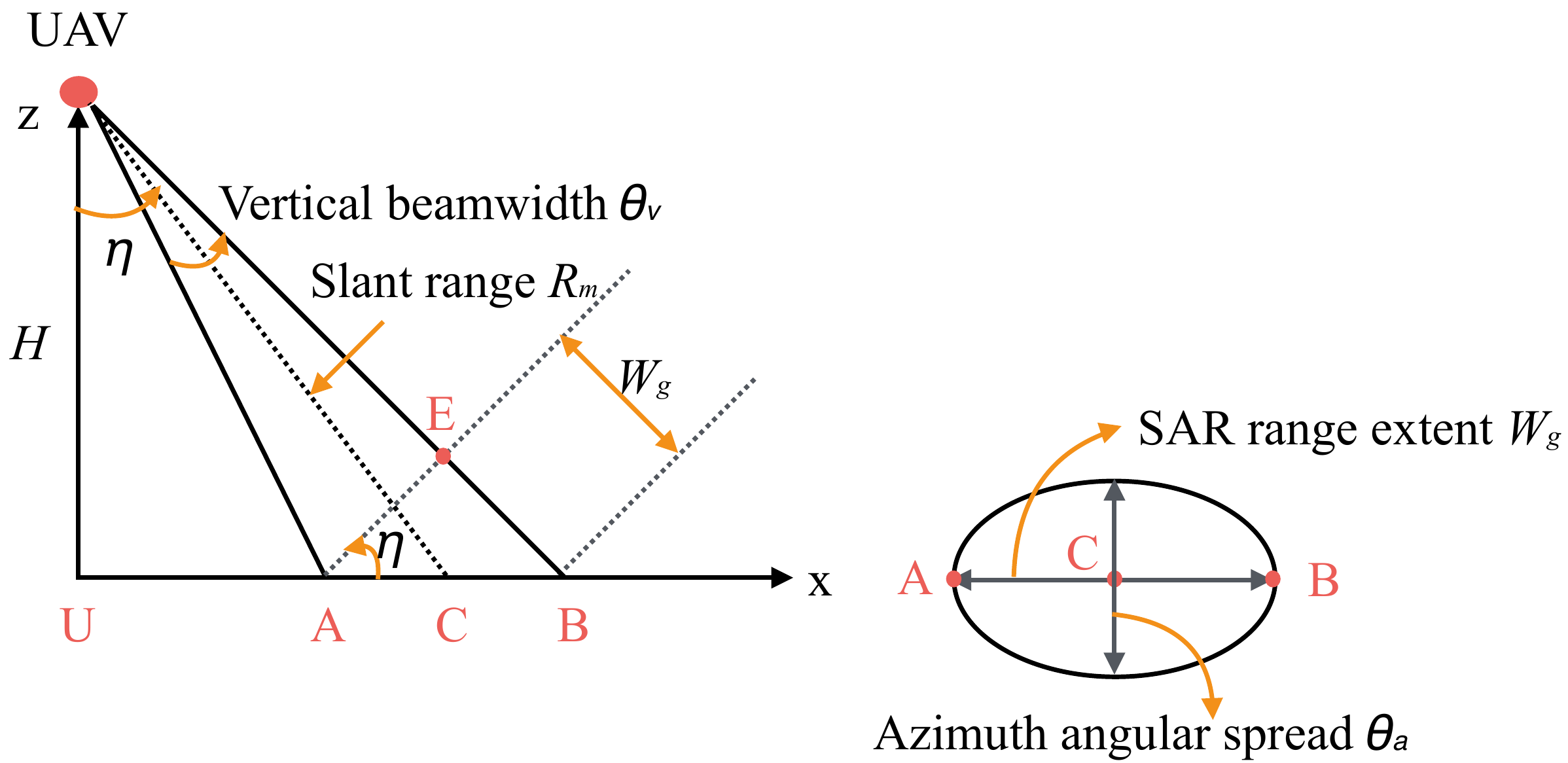}
\caption{The $(x,z)$-plane view of the side-looking BS-UAV bistatic SAR effective sensing area, where the UAV heading is perpendicular to the paper,
and $W_g$ is the SAR range extent.}
\label{sensing}
\end{figure}

Considering the characteristics of the side-looking SAR, the cosine of the UAV elevation angle is
\begin{equation}
\cos \eta \approx \frac{L_{\rm AE}}{W_g} \approx \frac{\theta_v R_m}{W_g} \approx \frac{\lambda_c R_m}{W_a W_g},
\end{equation}
where $R_m$ is the slant range (distance) from the SAR antenna to the center of $\mathcal U_t$ (i.e., point $C$);
$W_a$ is the height of the SAR antenna,
and $W_g$ is the SAR range extent, i.e., the width of the ground swath covered by the SAR beam, to the right of the UAV;
and the vertical beamwidth of the SAR is $\theta_v = \lambda_c / W_a$.
The azimuth angular spread is $\theta_a = \lambda_c / L_a$,
with $L_a$ being the antenna length parallel to the direction of the UAV's heading.
This spread is due to the interference of the waves emitted from and received by the dipoles of the antenna~\cite{collin85}.

{\blue The effective sensing area is ellipsoidal, denoted by $\mathcal U_t$.
The ellipsoidal effective sensing area is significantly smaller than the coverage area of the BS, $\mathcal A$, and is used to sample different parts of the coverage area for object detection.
The major and minor axes of the ellipsoidal sensing area, $\mathcal U_t$, are
$W_g=\frac{\lambda_c R_m}{W_a \cos \eta}$ and $\theta_a = \lambda_c / L_a$, respectively.}
We can obtain from Fig.~\ref{sensing} that $R_m = H/ \cos \eta$.
Therefore, at any moment, the size of the effective sensing area $\mathcal U_t$, i.e., $S$, is
\begin{equation}\label{eq.size}
S = \frac{\pi}{4} \times \frac{\lambda_c}{L_a} \times \frac{\lambda_c H}{W_a \cos^2 \eta}.
\end{equation}

{\blue Given the range and azimuth resolutions of the SAR, and the center and size of the effective sensing area,
the BS-UAV bistatic SAR can efficiently sense, detect, and distinguish ground objects in the disk region while the UAV is flying.
The details are provided in Section III.}

\section{Proposed Framework for UAV-assisted Joint Communication and Radar Sensing}\label{sec.prob}
The UAV performs sensing at a series of landmarks at specified time.
The BS-UAV bistatic SAR satisfies the resolution requirements to distinguish objects.
The 2D coordinates of the landmark to be sensed at the $t$-th time slot are $\mathbf q_g^t=[x_g^t, y_g^t]^T,~\forall t$.
We have $G \in \mathcal U_t,~\forall t$.

Through antenna configuration, the ellipsoidal shape of $\mathcal U_t$ can be approximated by a circular area,
i.e., $\theta_a = W_g$.
The landmark of interest is inside the sensing area $\mathcal U_t$.
Therefore, the distance between the landmark $G$ and the center of the sensing area $\mathcal U_t$, i.e., point $C$,
must not exceed the radius of $\mathcal U_t$ at any time slot $t$.
With the coordinates of the center $C$ and the diameter of $\mathcal U_t$, i.e., $\lambda_c/ L_a$, we have
\begin{equation}
\|\mathbf q_g^t- \mathbf q_c^t \| \le \frac{\lambda_c}{2 L_a},~\forall t.
\end{equation}

The UAV flies at a constant altitude $H$, and detects objects.
The UAV minimizes the overall propulsion energy consumption during the considered time horizon,
while maintaining acceptable sensing resolution by reusing cellular communication signals.
This can be formulated as
\begin{subequations}\label{p2}
\begin{align}
&\min_{\{\mathbf q_t, V_t, \forall t\}} \sum_{t=1}^{T_w} P_t \delta   \label{p2.0}\\
&\text{s.t.~} \frac{c}{B (\sin \eta + \sin \theta_t)} \le d_{\min}, ~\forall t, \label{p2.1} \\
& \quad~~  \frac{\lambda_c H}{T_d V_t \cos \eta} \le d_{\min} , ~\forall t, \label{p2.2} \\
& \quad~~ \|\mathbf q_t- \mathbf q_{t-1} \|  \leq \delta V_t, ~\forall t, \label{p2.3} \\
& \quad~~  0 \le V_t \le V_m,~\forall t, \label{p2.4} \\
& \quad~~  \|\mathbf q_g^t- \mathbf q_c^t \| \le \frac{\lambda_c}{2L_a} ,~\forall t, \label{p2.5}
\end{align}
\end{subequations}
where constraints \eqref{p2.1} and \eqref{p2.2} are the range and azimuth resolution requirements, respectively;
\eqref{p2.3} and \eqref{p2.4} are the constraints on the UAV's mobility and maximum speed, respectively;
and \eqref{p2.5} guarantees that the landmarks of interest are always captured at given time slots.

Problem \eqref{p2} is non-convex because of the non-convexity of the objective \eqref{p2.0} and constraints
\eqref{p2.1} and \eqref{p2.5}.
In what follows, we employ the SCA and BCD techniques to convexify \eqref{p2}, and attain a quality trajectory.


\subsection{SCA-based Convexification}

The convexification starts with the non-convex term of $P_t$ in the objective \eqref{p2.0}, i.e.,
$\left(\sqrt{1+\frac{{V_t}^4}{4v_0^4}}-\frac{{V_t}^2}{2 v_0^2}\right)^{\frac{1}{2}}$ in \eqref{speed},
by introducing new auxiliary variables $\{q_t \geq 0,\, \forall t =1, \cdots, T_w\}$:
\begin{equation}
q_t^2 = \sqrt{1+\frac{{V_t}^4}{4v_0^4}}-\frac{{V_t}^2}{2 v_0^2}, ~\forall t,
\end{equation}
which can be reorganized as
\begin{equation}\label{mu}
\frac{1}{q_t^2} = q_t^2 + \frac{{V_t}^2}{v_0^2}, ~\forall t.
\end{equation}
Herein, the second component on the right-hand side (RHS) of \eqref{speed} can be replaced by a linear element $P_1 q_t$,
with a newly added constraint \eqref{mu}.
As such, $P_t$ is transformed into
\begin{equation}\label{newpm}
{\tilde P}_t := P_0 + \frac{3P_0}{U_{tip}^2} {V_t}^2 + P_1 q_t + \frac{d_f}{2} \rho s A {V_t}^3, ~\forall t.
\end{equation}
Here, ${\tilde P}_t$ is jointly convex in $(V_t, q_t)$.

With a fixed $\delta$, problem \eqref{p2} becomes
\begin{subequations}\label{p3}
\begin{align}
&\min_{\{\mathbf q_t, V_t, q_t, \forall t \}} \sum_{t=1}^{T_w} \tilde P_t \label{p31}\\
\text {s.t.} ~& \frac{1}{q_t^2} \leq q_t^2 + \frac{{V_t}^2}{v_0^2 }, ~\forall t, \label{p32}\\
&\eqref{p2.1} - \eqref{p2.5}. \notag
\end{align}
\end{subequations}
Constraint \eqref{p32} is attained by slackening the equality in \eqref{mu} with inequality.
Problems \eqref{p2} and \eqref{p3} are equivalent.
The reason is that, if \eqref{p32} holds with inequality for any $t$,
one could always diminish the value of the relevant variable $q_t$ to save the overall power until \eqref{p32} takes equality \cite{zyong}.

Problem \eqref{p3} is still non-convex due to the non-convex constraint \eqref{p32},
and nevertheless can be resolved with the SCA approach~\cite{sca18}
through determining the global lower limit of \eqref{p32} at a specified local point.
Specifically, the left-hand side (LHS) of \eqref{p32} exhibits convexity in $q_t$, and the RHS exhibits convexity
in $\{q_t, V_t\}$.
Because the first-order Taylor expansion acts as a global lower limit of a convex function \cite{Boyd},
we can attain the lower limit for the RHS of \eqref{p32}, as given by:
\begin{equation}\label{eq.q}
\begin{aligned}
q_t^2 + \frac{{V_t}^2}{v_0^2} \geq &\, q_t^{(\ell)2} + \frac{1}{v_0^2} {V_t^{(\ell)}}^2 + 2q_t^{(\ell)} \left( q_t - q_t^{(\ell)} \right)  \\
& +\frac{2}{v_0^2} V_t^{(\ell)} \left( V_t - V_t^{(\ell)} \right),
\end{aligned}
\end{equation}
where $q_t^{(\ell)}$ and $V_t^{(\ell)}$ are the respective values of the variables at the $\ell$-th iteration of the SCA method.

We proceed to convexify constraints \eqref{p2.1} and \eqref{p2.5}, since the other constraints \eqref{p2.2}-\eqref{p2.4} are all convex by now.
This starts by rewriting \eqref{p2.1} as 
\begin{equation}
\sin \theta_t \ge \frac{c}{B d_{\min}} - \sin \eta.
\end{equation}
With $\sin^2 \theta_t = 1- \cos^2 \theta_t$, we have
\begin{equation}\label{eq.range2}
\frac{H_b^2}{\|\mathbf q_c^t- \mathbf q_b \| ^2+H_b^2} \le 1-  \left( \frac{c}{B d_{\min}} - \sin \eta \right)^2.
\end{equation}
Introduce slack variables $w_t=\sin \alpha_t$ and $u_t = \cos \alpha_t, \forall t$.
From \eqref{eq.mob} and \eqref{eq.sincos}, we have additional constraints on $w_t$ and $u_t$:
\begin{subequations}\label{eq.uvwt}
\begin{align}
& -1 \le w_t \le \frac{y_t - y_{t-1}}{\delta V_t}, ~\forall t, \label{eq.wt}\\
& \frac{x_t - x_{t-1}}{\delta V_t} \le u_t \le 1, ~\forall t. \label{eq.ut}
\end{align}
\end{subequations}
Constraint \eqref{eq.uvwt} is non-convex because of the coupling between $V_t$ and $w_t$ (or $u_t$).

With $w_t$ and $u_t$, \eqref{eq.range2} can be reorganized as
\begin{equation}\label{eq.explicit}
\begin{aligned}
&(x_t + w_t H \tan \eta -x_b)^2+(y_t - u_t H \tan \eta -y_b)^2  \\
& \ge \frac{H_b^2}{1-  \left( \frac{c}{B d_{\min}} - \sin \eta \right)^2}-H_b^2 :=C_r,~\forall t.
\end{aligned}
\end{equation}
The LHS of \eqref{eq.explicit} is convex in $\{\mathbf q_t, w_t, u_t\}$.
The RHS of \eqref{eq.explicit} is a constant and defined as $C_r$ for illustration convenience,
given the predetermined parameters $H_b,~c,~B,~d_{\min}$, and $\eta$.
We convexify \eqref{eq.explicit} by determining the first-order Taylor expansion of its LHS at the local point attained
during the $\ell$-th iteration of the SCA:

Given $\{w_t^{(\ell)}, u_t^{(\ell)}\}$, \eqref{eq.explicit} can be linearized in $\mathbf q_t$:
\begin{equation}\label{eq.explicit2}
\begin{aligned}
& (x_t^{(\ell)} + w_t^{(\ell)} H \tan \eta -x_b)^2 + (y_t^{(\ell)} - u_t^{(\ell)} H \tan \eta -y_b)^2 \\
& + 2(x_t^{(\ell)} + w_t^{(\ell)} H \tan \eta -x_b)(x_t-x_t^{(\ell)})  \\
& + 2(y_t^{(\ell)} - u_t^{(\ell)} H \tan \eta -y_b)(y_t-y_t^{(\ell)}) \ge C_r,~\forall t.
\end{aligned}
\end{equation}

Given $\{\mathbf q_t^{(\ell)}, V_t^{(\ell)}, q_t^{(\ell)}\}$, \eqref{eq.explicit} can be linearized in $\{ w_t, u_t \}$:
\begin{equation}
\begin{aligned}
& (x_t^{(\ell)} + w_t^{(\ell)} H \tan \eta -x_b)^2 + (y_t^{(\ell)} - u_t^{(\ell)} H \tan \eta -y_b)^2 \\
& + H \tan \eta (x_t^{(\ell)} + w_t^{(\ell)} H \tan \eta -x_b)(w_t-w_t^{(\ell)})  \\
& + H \tan \eta (y_t^{(\ell)} - u_t^{(\ell)} H \tan \eta -y_b)(u_t-u_t^{(\ell)}) \ge C_r, \forall t. \label{eq.explicit3}
\end{aligned}
\end{equation}

As done to \eqref{eq.explicit}, constraint \eqref{p2.5} can be rewritten as
\begin{equation}\label{eq.tagcircle}
(x_t + w_t H \tan \eta -x_g^t)^2+(y_t - u_t H \tan \eta -y_g^t)^2 \le \left( \frac{\lambda_c}{2L_a} \right)^2, \forall t,
\end{equation}
which is convex in $\{\mathbf q_t, w_t, u_t\}$.

As a result, problem \eqref{p3} is transformed into
\begin{subequations}\label{p4}
\begin{align}
&\min_{\{\mathbf q_t, V_t, q_t,w_t,u_t, \forall t \}} \sum_{t=1}^{T_w} \tilde P_t \label{p41}\\
\text {s.t.} ~& \frac{1}{q_t^2} \leq q_t^{(\ell)2} + \frac{1}{v_0^2} {V_t^{(\ell)}}^2 + 2q_t^{(\ell)} \left( q_t - q_t^{(\ell)} \right)  \notag \\
& \qquad ~ +\frac{2}{v_0^2} V_t^{(\ell)} \left( V_t - V_t^{(\ell)} \right), ~\forall t, \label{p42}\\
&\eqref{p2.2} - \eqref{p2.4}, ~\eqref{eq.uvwt},~ \eqref{eq.explicit2}-\eqref{eq.tagcircle}. \notag
\end{align}
\end{subequations}
where \eqref{p42} tightens the original constraint \eqref{p32} by using a lower bound of its RHS.
Since problem \eqref{p4} is a tightened version of problem \eqref{p3},
the feasible region of \eqref{p4} is also the feasible region of \eqref{p3};
not the other way around.

\subsection{Block Coordinate Descent (BCD)}
Problem \eqref{p4} is still non-convex due to the non-convexity of constraint \eqref{eq.uvwt}.
We apply the BCD to optimize $\{\mathbf q_t, V_t, q_t\}$ and $\{w_t, u_t\}$ in an alternating manner,
because the variables $V_t$ and $\{w_t, u_t\}$ are coupled in the non-convex constraint \eqref{eq.uvwt}:

Given fixed $\{w_t, u_t\}$, problem \eqref{p4} is reduced to the following convex problem:
\begin{subequations}\label{p5}
\begin{align}
&\min_{\{\mathbf q_t, V_t, q_t, \forall t \}} \sum_{t=1}^{T_w} \tilde P_t \label{p51}\\
\text {s.t.} ~& w_t^{(\ell)} \le \frac{y_t - y_{t-1}}{\delta V_t}, ~\forall t, \label{p52}\\
& \frac{x_t - x_{t-1}}{\delta V_t} \le u_t^{(\ell)}, ~\forall t, \label{p53} \\
& (x_t + w_t^{(\ell)} H \tan \eta -x_g^t)^2+(y_t - u_t^{(\ell)} H \tan \eta -y_g^t)^2 \notag \\
& \le \left( \frac{\lambda_c}{2L_a} \right)^2, \forall t, \\
&\eqref{p2.2} - \eqref{p2.4}, ~\eqref{eq.explicit2}, ~\eqref{p42}. \notag
\end{align}
\end{subequations}

Given fixed $\{\mathbf q_t, V_t, q_t\}$, problem \eqref{p4} becomes a feasibility checking problem
to find the values of $\{w_t, u_t\}$ that satisfy the following convex constraints:
\begin{subequations}\label{p6}
\begin{align}
&\text{find} ~\{w_t, u_t, \forall t \} \label{p61}\\
\text {s.t.} ~& -1 \le w_t \le \frac{y_t^{(\ell)} - y_{t-1}^{(\ell)}}{\delta V_t^{(\ell)}}, ~\forall t, \label{p62}\\
& \frac{x_t^{(\ell)} - x_{t-1}^{(\ell)}}{\delta V_t^{(\ell)}} \le u_t \le 1, ~\forall t, \label{p63} \\
& (x_t^{(\ell)} + w_t H \tan \eta -x_g^t)^2+(y_t^{(\ell)} - u_t H \tan \eta -y_g^t)^2 \notag \\
& \le \left( \frac{\lambda_c}{2L_a} \right)^2, ~\forall t, \\
& \eqref{eq.explicit3}. \notag
\end{align}
\end{subequations}
Problems \eqref{p5} and \eqref{p6} can be solved by existing convex tools, such as MATLAB CVX.

{\blue When considering 3D trajectory planning, we can potentially decouple the optimizations of the 2D horizontal trajectory and the altitude by using alternating optimization techniques. Given fixed altitudes of the UAV $H_t,\, \forall t$, the 2D trajectory can be designed using the proposed Algorithm~1. Given fixed 2D trajectory (and auxiliary variables), i.e., $\{\mathbf q_t, V_t, q_t,w_t,u_t, \forall t\}$, the varying UAV altitude $H_t, \forall t$, can be potentially optimized by solving a similar subproblem to \eqref{p6} by replacing $w_t$ and $u_t$ with $H_t$ in (29a), and $H$ with $H_t$ in (29d). The subproblem can be convexified and solved in the same way as problem \eqref{p6}, as described in Algorithm 1. The horizontal trajectory and the altitude can be optimized in an alternating manner until convergence. The detailed design and validation of the 3D trajectory planning will be presented in our subsequent study.}

\begin{algorithm}[t]
\caption{The proposed trajectory planning algorithm for the cellular-aided radar sensing to problem \eqref{p2}.}
\label{algo.sensing}
\begin{algorithmic}[1]
\State {\bf Initialization:} Generate a feasible initial flight path for the UAV, i.e.,
$\{ \mathbf q_{t}^{(0)}, V_{t}^{(0)}, q_{t}^{(0)}, t=1,\cdots,T_w\}$,
and input initial values for the slack variables $\{w_t^{(0)}, u_{t}^{(0)}, t=1,\cdots,T_w\}$.
\For {$\ell$ = 1, 2, ...}
\State Approximate the propulsion power $P_t$ by \eqref{newpm}, \eqref{p32},
and \eqref{eq.q} by the SCA technique.
\State Introduce slack variables $w_t$ and $u_t, \forall t$, and bound them by constraint \eqref{eq.uvwt}.
\State Convexify constraint \eqref{p2.1} by \eqref{eq.explicit2} and \eqref{eq.explicit3},
and rewrite constraint \eqref{p2.5} into a convex one \eqref{eq.tagcircle}.
\State Given fixed $\{w_t^{(\ell-1)}, u_t^{(\ell-1)}\}$, solve problem \eqref{p5} by the interior point method to update the optimization variables
$\{ \mathbf q_{t}^{(\ell)}, V_{t}^{(\ell)}, q_{t}^{(\ell)} \}$ to design the UAV trajectory.
\State Given $\{ \mathbf q_{t}^{(\ell)}, V_{t}^{(\ell)}, q_{t}^{(\ell)} \}$, solve problem \eqref{p6} by the interior point method to update the slack
variables $\{w_t^{(\ell-1)}, u_t^{(\ell-1)}\}$.
\State Update $\ell \leftarrow \ell+1$.
\EndFor
\end{algorithmic}
\end{algorithm}

\subsection{Algorithm and Complexity}
Algorithm~\ref{algo.sensing} summarizes the steps proposed in this section to convexify and solve efficiently problem \eqref{p2}.
The algorithm is computationally dominated by Steps 6 and 7,
as Steps 3 to 5 compute linear functions with a complexity of $\mathcal{O}(N)$.
{\blue After a series of mathematical manipulations and transformations, the original problem \eqref{p2} is recast as \eqref{p4} with a convex objective and non-convex constraints, and further decomposed into two subproblems, i.e., problems \eqref{p5} and \eqref{p6}, to convexify the non-convex constraints.
Problems \eqref{p5} and \eqref{p6} are convex programs, given their convex objective functions and constraints (i.e., all the constraints are either linear or quadratic).
In this sense, Steps 6 and 7 each solve a convex program and update the optimization variables, i.e., by running the interior point method with the complexity of $\mathcal O(N^{3.5})$ per iteration~\cite{Nemirovskii}.
The overall computational complexity of Algorithm~\ref{algo.sensing} is $\mathcal{O}(N^{3.5})$ per iteration.
Interested readers can refer to~\cite{Boyd} for a detailed introduction of 
the interior point method and convex optimization in general.


Let $\mathbf x$ and $\mathbf x^*$ denote a feasible solution and the optimal solution to problem~\eqref{p4}, respectively. Also, let $P$ denote the objective value of problem~\eqref{p4}.
As established in~\cite{complexity}, given the convergence precision $\epsilon$ of the algorithm, i.e., $| P(\mathbf x) - P(\mathbf x^*)| \le \epsilon$, the interior-point method takes $\mathcal{O}(\log\frac{1}{\epsilon})$ iterations before convergence.
Therefore, 
Algorithm~\ref{algo.sensing} has a polynomial complexity of $\mathcal{O}(N^{3.5}\log\frac{1}{\epsilon})$.}

\begin{table}[t]
\renewcommand{\arraystretch}{1.3}
\caption{The Parameters of the BS-UAV bistatic SAR \cite{zyong, moccia11}}
\begin{center}
\begin{tabular}{l | l} 
\hline
\text{Parameter}    &\text{Value} \\ \hline
Transmission bandwidth, $B$                                  & 150 MHz     \\ \hline
Wavelength, $\lambda_c$                                        & 0.1 m       \\ \hline
Coherent integration time, $T_d$                            & 1 s              \\ \hline
SAR observation angle, $\eta$                                 & $\pi$/4        \\ \hline
Minimum object distance, $d_{\min}$                    & 20 m             \\ \hline
UAV weight and altitude, $H$                                   &2 kg, 1000 m      \\ \hline
Maximum UAV speed, $V_m$                                  & 50 m/s            \\ \hline
\text{Blade profile power and tip speed, $P_0$ and $U_{tip}$}     &3.4 W, 60 m/s  \\ \hline
Rotor induced power and velocity, $P_1$ and $v_0$              & 118 W, 5.4 m/s  \\ \hline
\text{Rotor solidity and disc area, $s$ and $A$}                       &0.02, 0.5 m$^2$\\ \hline
\text{Air density and fuselage drag fraction, $\rho$ and $d_f$}            &1.225 kg/m$^3$, 0.3 \\ \hline
\end{tabular}
\end{center}
\label{tab.rotary}
\end{table}

\begin{figure}[t]
\centering
\includegraphics[width=0.5\textwidth]{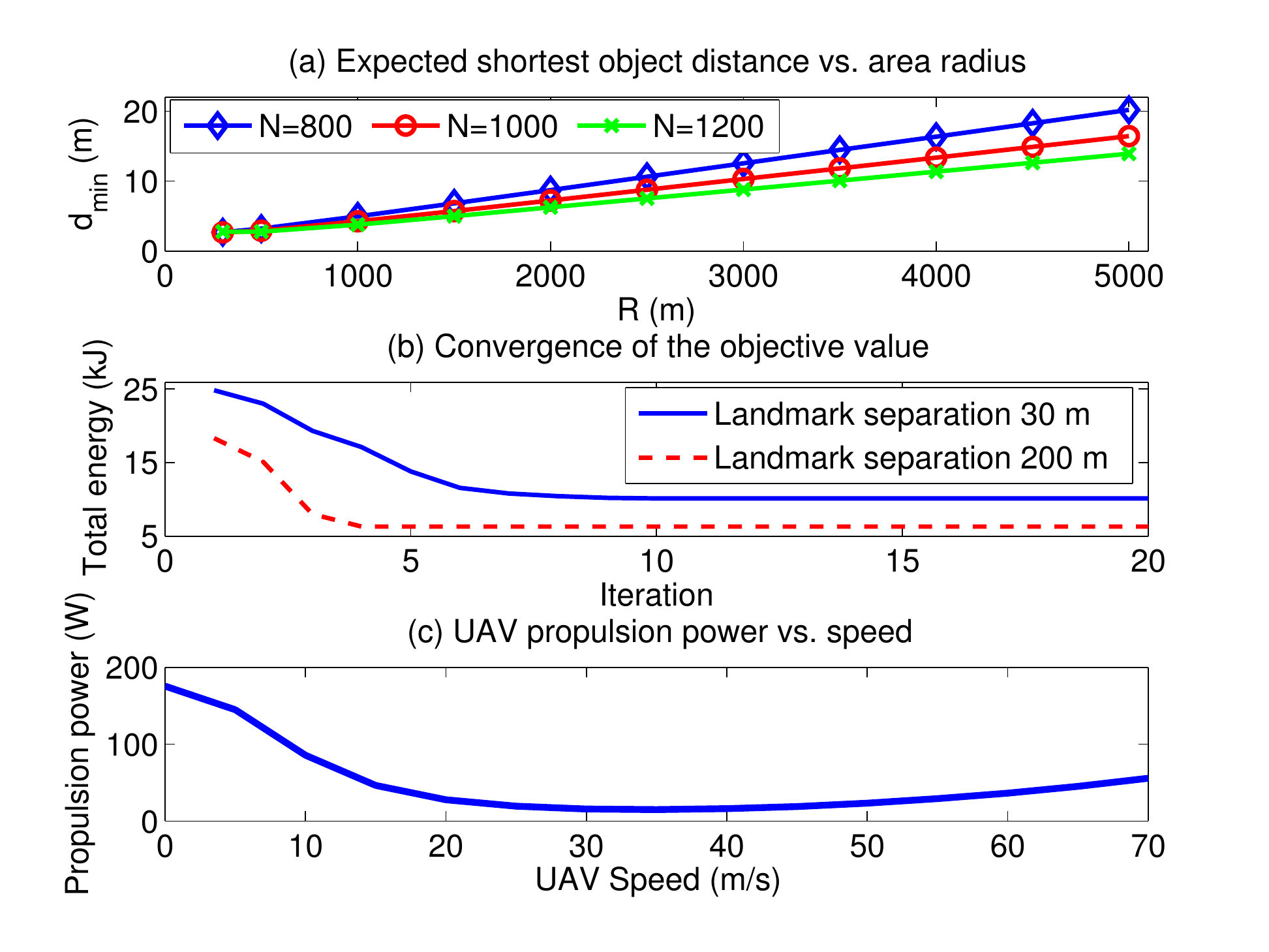}
\caption{Fig.~\ref{fig41}(a) plots the expected shortest distance between any two objects in the area, $d_{\min}$,
vs. the radius of the given area, $R$;
Fig.~\ref{fig41}(b) plots the convergence of the objective value \eqref{p41} under the default setting;
and Fig.~\ref{fig41}(c) plots the UAV propulsion power vs. speed.}
\label{fig41}
\end{figure}

\begin{figure}[t]
\centering
\includegraphics[width=0.5\textwidth]{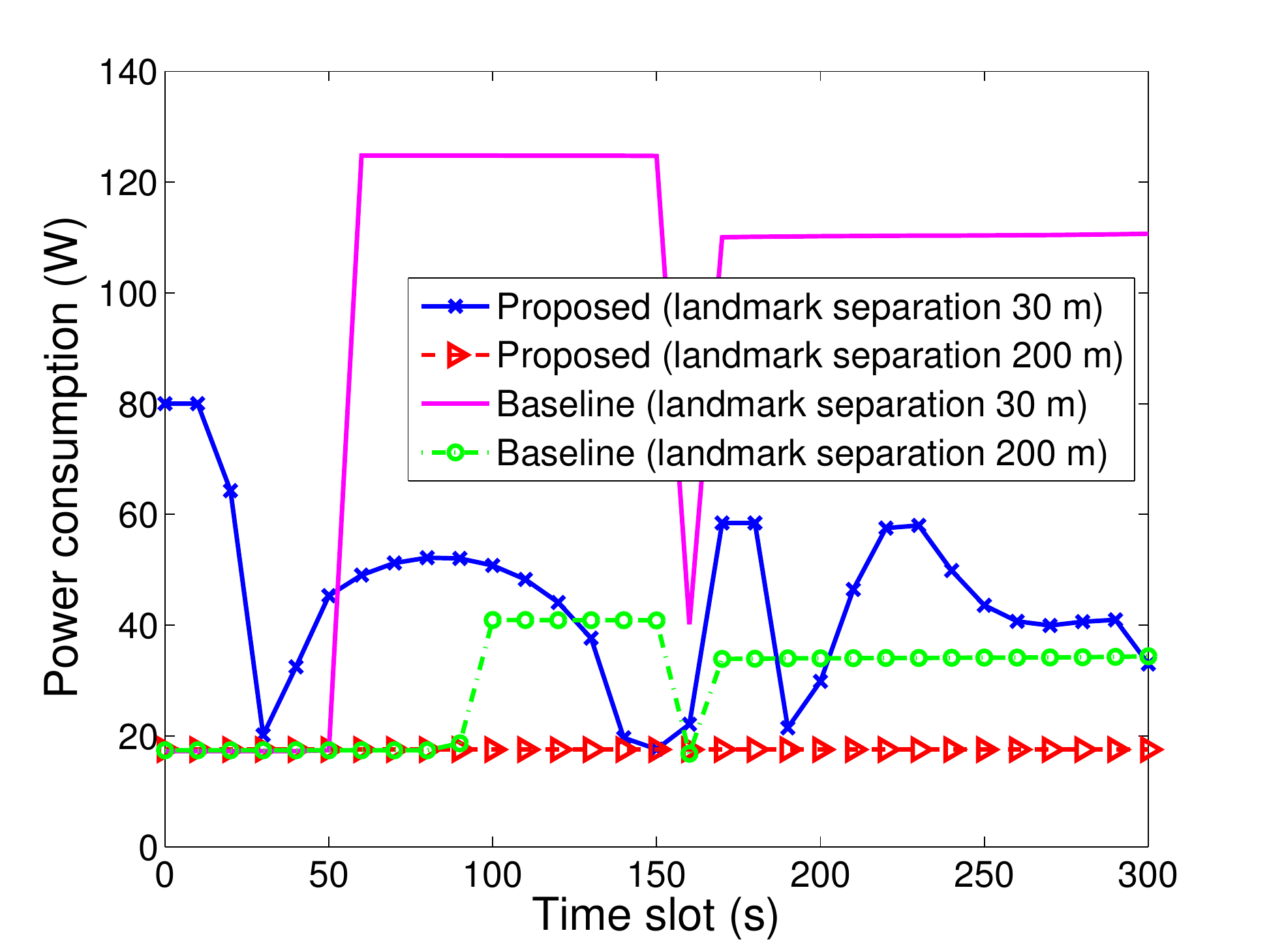}
\caption{The UAV per-slot energy consumption by the proposed and baseline schemes when the landmarks of interest are separated
by $30$ m and $200$ m, under the default setting.}
\label{energy}
\end{figure}

\begin{figure}[t]
\centering
\includegraphics[width=0.5\textwidth]{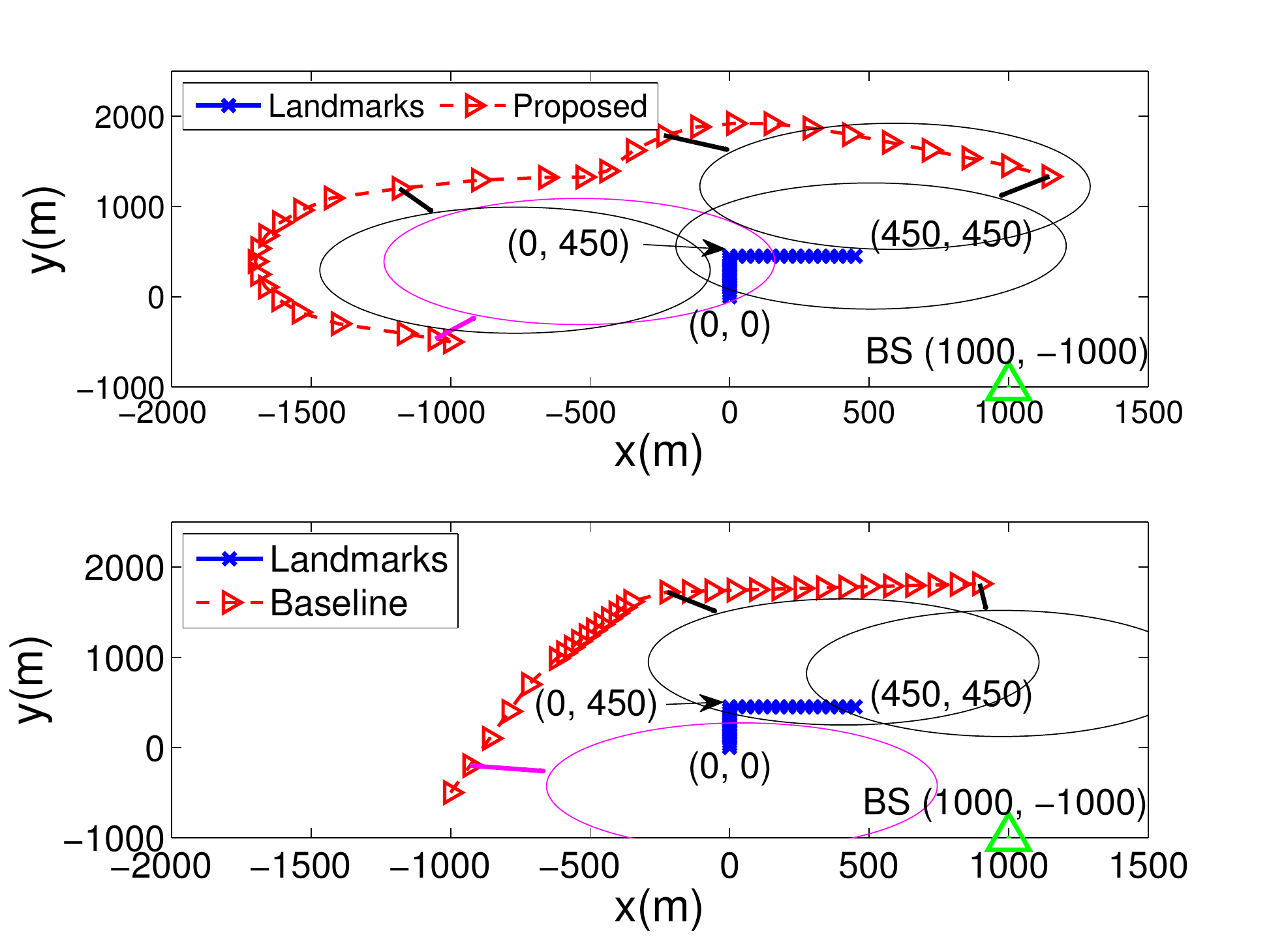}
\caption{The UAV trajectory by the proposed and baseline schemes,
where the landmarks of interest are separated by $30$ m, and $H=1000$ m.}
\label{figfigturn3}
\end{figure}

\begin{figure}[t]
\centering
\includegraphics[width=0.5\textwidth]{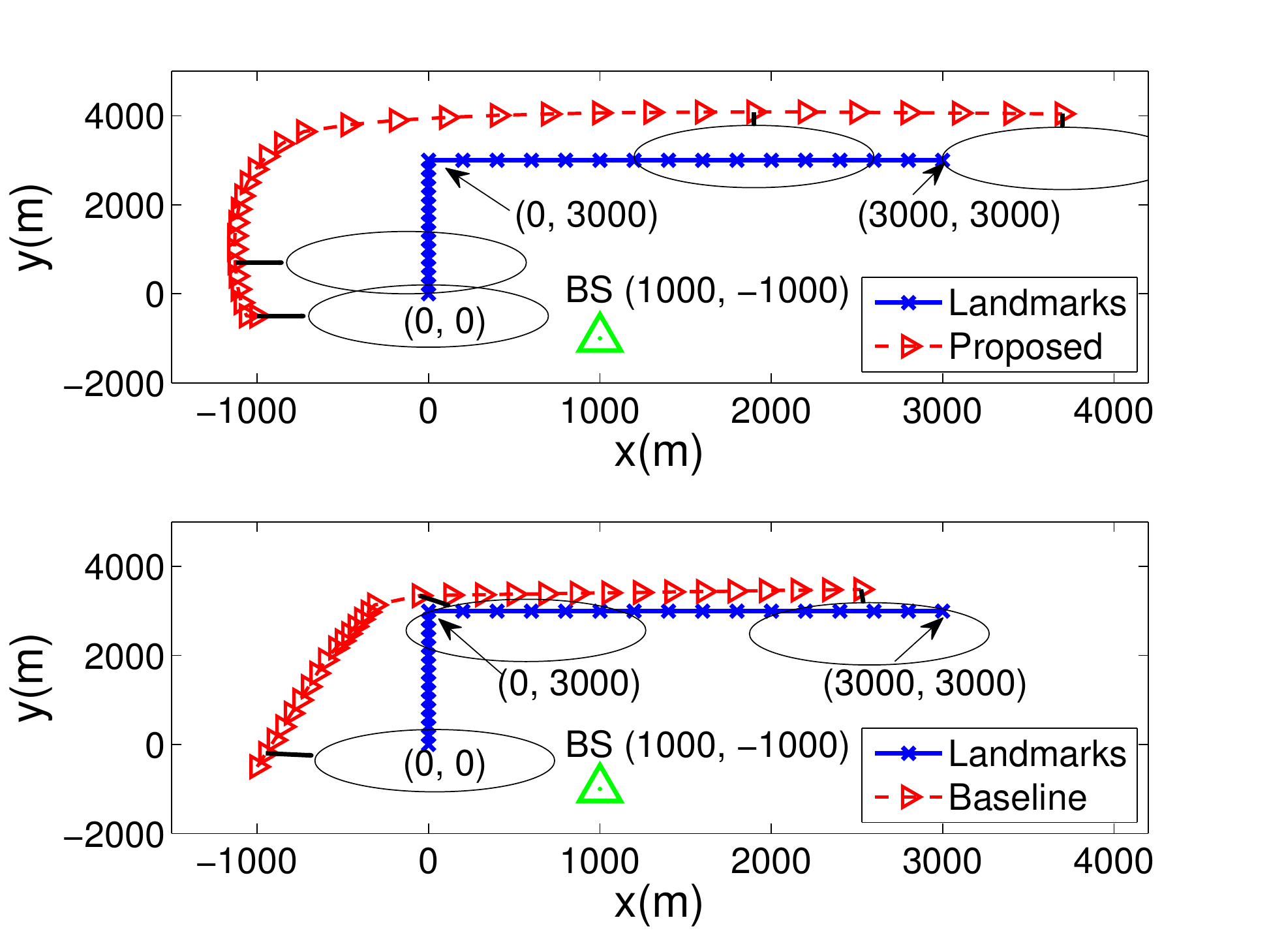}
\caption{The UAV trajectory by the proposed and baseline schemes,
where the landmarks of interest are separated by $200$ m, and $H=1000$ m.}
\label{figfigturn20}
\end{figure}

\section{Simulation Results}\label{sec.sim}
This section provides the simulation results of the proposed trajectory planning algorithm for BS-UAV bistatic SAR using MATLAB.
The entire sensing period lasts $T=300$ s with each time slot of $\delta = 0.5$~s, unless stated otherwise.
The default setting of the landmarks (and associated sensing time) is a series of points with the first $15$ of them,
i.e., when $t \le 150$ s, along the $y$-axis starting from $(0,0)$,
and the rest along the $x$-axis starting from $\mathbf q_g^{16}$.
The adjacent landmarks are apart for an equal distance.
The UAV initial location is $\mathbf q_0=[-1000, -500]^T$~m.
The BS is located at $\mathbf q_b=[1000, -1000]^T$~m, with the height $H_b= 50$ m.
The other parameters concerning the BS-UAV bistatic SAR sensing performance and the UAV propulsion power are provided
in Table~\ref{tab.rotary}.

{\blue 
We note that the standard sub-6 GHz band has been extensively deployed to provide broad 5G coverage, where the antennas are typically horizontally or quasi-horizontally oriented in the sub-6 GHz band~\cite{cudak}. 
The transmission bandwidth is typically narrow in the sub-6 GHz band, e.g., up to 100 MHz, and consequently the range resolution could be poor~\cite{gppprotocol}. Nevertheless, a bandwidth of 150 MHz can be achieved in a standard sub-6 GHz band by using carrier aggregation techniques~\cite{khan}. For instance, China Mobile deployed a 4G and 5G concurrent integrated network supporting a transmission bandwidth of up to 160 MHz at the 2.6 GHz band in 2019~\cite{huaweicell}. On the other hand, the 5G new radio (NR) has also specified the use of mmWave bands. The propagation of mmWave signals can be limited within a close range of the BS, subject to antenna orientation (i.e., downtilt)~\cite{cudak}. Nevertheless, the signals can still be reflected by objects (e.g., with smooth metallic surfaces) or the edges of objects~\cite{Goddemeier, Jaeckel}, and 
captured by the UAV.
Given the quasi-optical property of mmWave signals~\cite{Giordani}, the ground reflections (and reflections by other non-smooth surfaces, e.g., walls, vegetation, etc.) are expected to be substantially weaker,
resulting in strong contrast to manifest objects~\cite{Giordani}.}

{\blue We also note that no existing studies have addressed the problem of energy-efficient trajectory design for the considered UAV-based bistatic SAR system, as discussed in Section~\ref{sec: related work}. In other words, no existing algorithm is directly comparable to the proposed algorithm. 
With due diligence, we come up with a new baseline scheme for the considered problem, which minimizes the total flight distance of the UAV without considering the energy consumption of the UAV,
i.e., $\min_{\mathbf q_t} \sum_t \| \mathbf q_t \|$.}
The baseline has a convex objective function, while its constraints can be convexified
in the same way as done in Algorithm~1.

Fig.~\ref{fig41}(a) shows the expected shortest distance between any two objects in the area, $d_{\min}$,
versus the radius of the given area $\mathcal A$,
$R$ (in meters), when there are $N=800,~1000$ or $1200$ objects in $\mathcal A$.
It can be seen that $d_{\min}$ increases with $R$, given the number of objects in the area;
i.e., the objects are more sparsely distributed when the area is larger.
When $R=5000$~m, the expected shortest distance between any two objects in the area $d_{\min}$ is about $20$~m.
Therefore, we choose the value of $d_{\min} = 20$ m.
%
Fig.~\ref{fig41}(b) depicts the convergence of the proposed scheme when the landmarks are separated by
$30$ m and $200$ m under two settings.
It can be seen that the objective function of problem \eqref{p4}, i.e., \eqref{p41}, can quickly converge within about only $20$ iterations.
When the two neighboring landmarks are apart for a longer distance,
the UAV consumes less energy and shows better energy efficiency.
This is because the UAV can move faster,
and the propulsion power first decreases and then increases at the increasing speed of the UAV,
as shown in Fig.~\ref{fig41}(c).
Fig.~\ref{fig41}(c) plots the UAV power consumption $P_t$ in \eqref{speed} by varying the instantaneous speed of a rotary-wing UAV
from $0$ m/s to $70$ m/s.
The UAV's propulsion power is independent of its trajectory and heading.
The UAV consumes the least power when its speed is about $35$ m/s, validating the results in Fig.~\ref{fig41}(b).

\begin{figure}[t]
\centering
\includegraphics[width=0.5\textwidth]{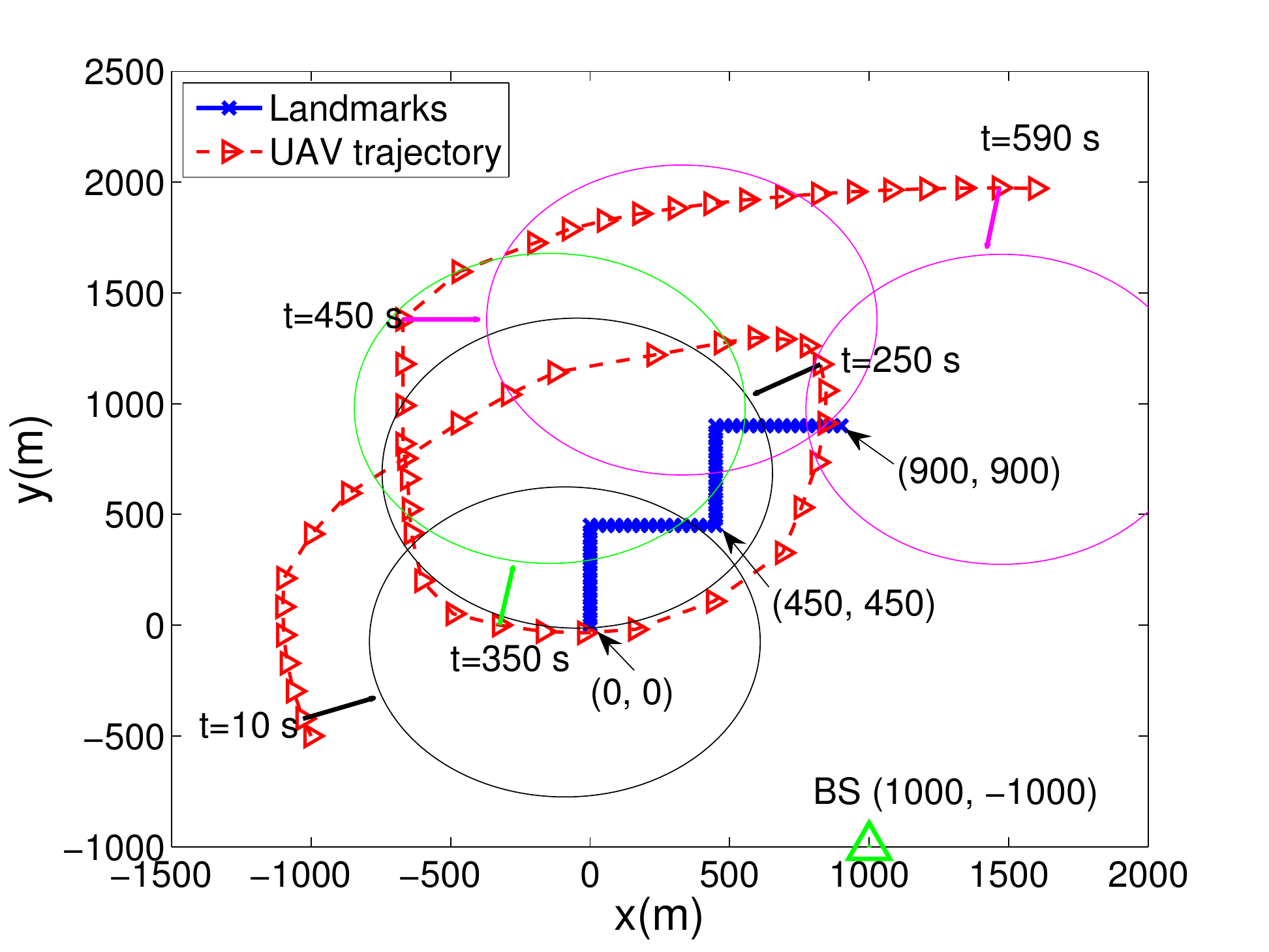}
\caption{The UAV trajectory by the proposed scheme, where the landmarks of interest are separated by $30$ m, $H=1000$ m, and $T=600$ s.}
\label{staircase}
\end{figure}

Fig.~\ref{energy} shows the per-slot energy consumption of the rotary-wing UAV conducting the BS-UAV bistatic SAR sensing under
the proposed and baseline schemes,
where the default setting of the landmarks is considered, i.e., with an equal distance of $30$~m or $200$~m
between adjacent landmarks.
It is revealed that the minimization of the flight distance does not necessarily lead to the minimization of the UAV's energy consumption. 
When the landmarks are separated by $30$ m, the proposed scheme can save $55\%$ of the total energy,
as compared to the baseline.

{\blue It is interesting to notice that the baseline scheme is better than the proposed algorithm in terms of per-slot UAV power consumption when $t \le 50$ s and the spacing between adjacent landmarks is $30$ m.
This is because the objective of the proposed algorithm, i.e., \eqref{p2.0},
is to minimize the total energy consumption.
As a result, the algorithm has the UAV fly slowly away from a landmark (while still keeping the landmark within its effective sensing area) at the beginning of the sensing mission to benefit the later stage of the mission in Fig.~\ref{energy}.
The UAV has to restrain its speed at the beginning of the mission at the cost of high energy consumption, 
as will be shown in Fig.~\ref{figfigturn3}.
On the other hand, the baseline scheme requires the UAV to fly fast towards the landmarks in the beginning and then slows down, to minimize the total flight distance to sense all landmarks.
As a result, the baseline scheme consumes less energy than the proposed algorithm at the beginning of the mission. Nevertheless, the proposed algorithm is much more energy-efficient when the entire mission is considered.}

%
%

\begin{figure}[t]
\centering
\includegraphics[width=0.5\textwidth]{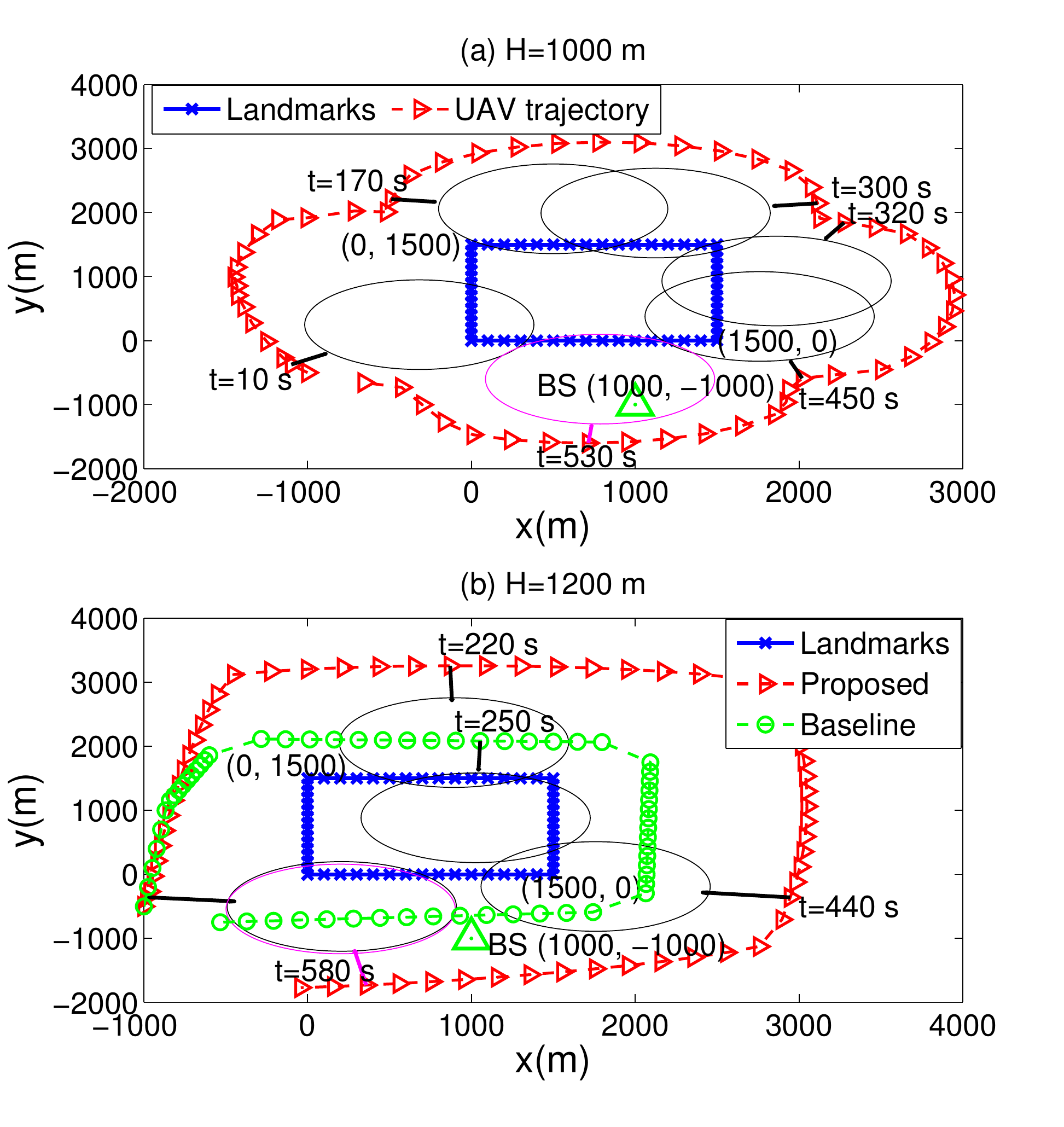}
\caption{The UAV trajectory by the proposed scheme, where the landmarks of interest are separated by $100$ m and $T=600$ s.}
\label{figsq21}
\end{figure}

\begin{figure}[t]
\centering
\includegraphics[width=0.5\textwidth]{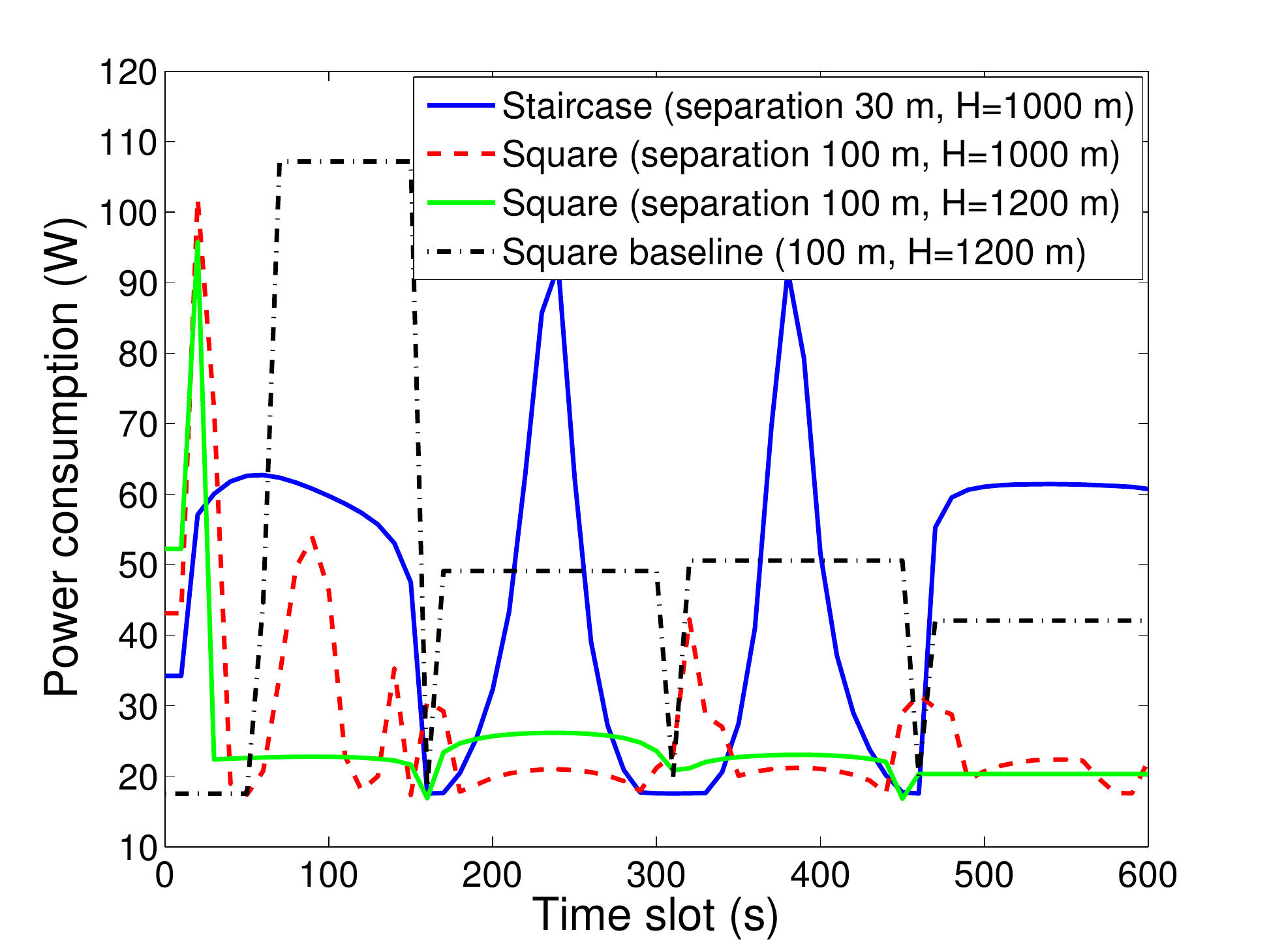}
\caption{The UAV per-slot energy consumption by the proposed scheme when the landmarks of interest are distributed like a staircase
(separated by $30$ m, $H=1000$ m), and a square (separated by $100$ m, $H=1000$ m and $1200$~m), and $T=600$ s.}
\label{engss}
\end{figure}

\begin{figure}[t]
\centering
\includegraphics[width=0.5\textwidth]{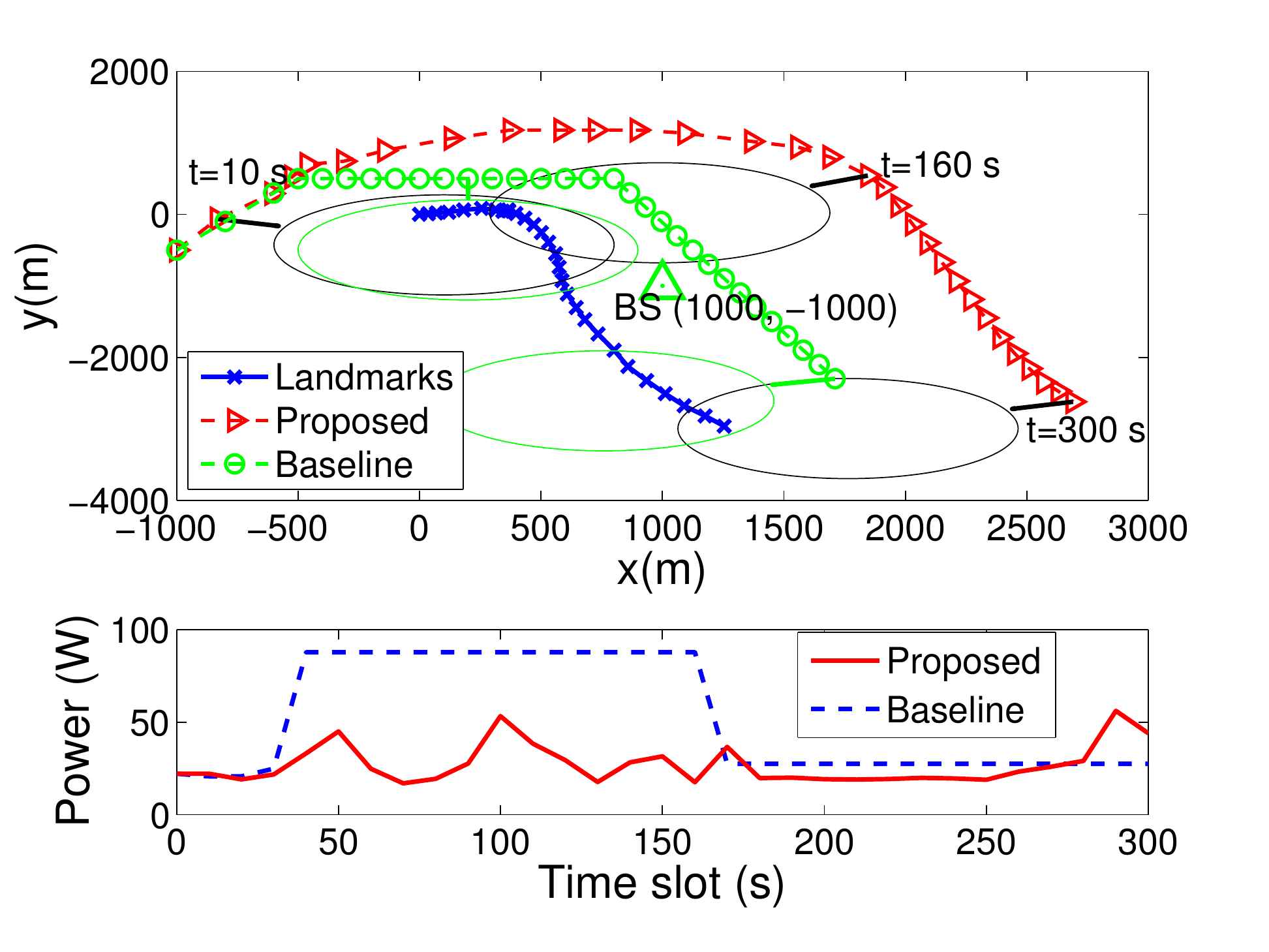}
\caption{The UAV trajectory and per-slot energy consumption by the proposed and baseline schemes when the landmarks of interest are distributed randomly, $H=1000$ m and $T=300$ s.}
\label{figrndtwo}
\end{figure}

Figs.~\ref{figfigturn3} and \ref{figfigturn20} plot the UAV trajectory (the red dash line with triangle markers)
and its sensing performance under the proposed and baseline schemes.
The black and magenta circles provide some examples of the effective sensing areas at different time slots.
The green triangle shows the location of the BS.
We see in Figs.~\ref{figfigturn3} and \ref{figfigturn20} that the UAV can always capture the landmarks within its sensing area at the time slots required.
Under the proposed schemes, the UAV seeks an energy-efficient flight path,
and is better adapted to the distribution of the landmarks.
In contrast, under the baseline schemes, the UAV first catches up with the nearest landmark at its highest speed,
and then flies at a relatively stable speed to satisfy the resolution requirements.
This can cause drastic changes in the instantaneous power consumption of the UAV under the baseline method,
as already shown in Fig.~\ref{energy}. 

Fig.~\ref{staircase} plots the UAV trajectory when $T=600$ s.
The landmarks are separated by $30$ m and distributed in a staircase (that is, two street blocks situated along a diagonal).
The effective sensing area of the BS-UAV bistatic SAR is indicated by the black, magenta, or green circles,
when $t=10$~s, $250$ s, $350$ s, $450$ s, and $590$ s, respectively.
It is shown in Fig.~\ref{staircase} that the BS-UAV bistatic SAR can always capture the landmarks in its sensing coverage
at the required time slots.

Fig.~\ref{figsq21} plots the UAV trajectory when the landmarks are separated by $100$ m
and distributed in a square (e.g., a closed loop around a triangular street block), $T=600$ s, and $H=1000$~m (Fig.~\ref{figsq21}(a))
or $1200$ m (Fig.~\ref{figsq21}(b)).
The effective sensing area of the BS-UAV bistatic SAR is shown
when $t=10$ s, $170$ s, $300$ s, $320$~s, $450$ s, and $530$ s, under the UAV elevation of $H=1000$ m,
and when $t=10$ s, $220$ s, $440$ s, and $580$ s, under the UAV elevation of $H=1200$~m.
It is observed that the UAV intends to fly around the landmarks for both effective sensing and energy saving.
The UAV trajectory of the corresponding baseline scheme is shown in Fig.~\ref{figsq21}(b) when $H=1200$ m;
see the green dashed line with circle markers.
The effective sensing area of the UAV is illustrated when $t=250$~s.
We see that the UAV takes a much shorter flight path around the landmarks to satisfy the sensing resolution requirements,
which is at the cost of a much higher propulsion energy consumption, as to be shown in Fig.~\ref{engss}.

\begin{figure}[t]
\centering
\includegraphics[width=0.5\textwidth]{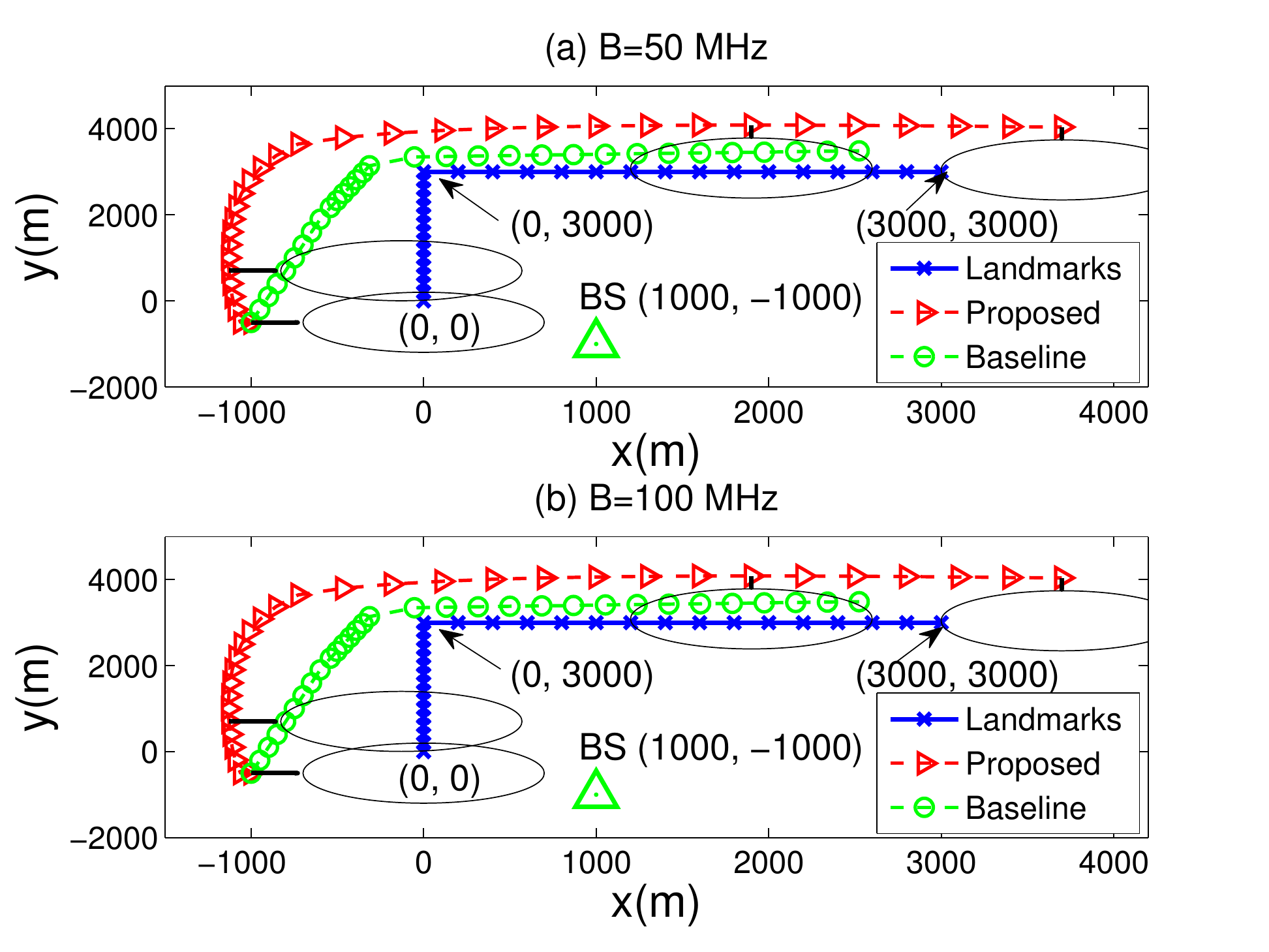}
\caption{{\blue The UAV trajectory by the proposed and baseline schemes when the landmarks of interest are separated
by $200$ m, and the transmission bandwidth is $50$ MHz and $100$ MHz under the default setting.}}
\label{figturnbd}
\end{figure}

\begin{table}[t]
\renewcommand{\arraystretch}{1.3}
\caption{{\blue The range resolutions of the BS-UAV bistatic SAR under different transmission bandwidths, where the transmitter
incident angle is $\theta_t = \pi/3$}}
\begin{center}
\begin{tabular}{| c| c| c| c| c|} 
\hline
Transmission bandwidth, $B$ (MHz)   & 50 &100 & 150 & 200   \\ \hline
Range resolution, $\delta_r$ (m)       &3.81 &1.91        &1.27                 & 0.95     \\ \hline
\end{tabular}
\end{center}
\label{tab.range}
\end{table}

Fig.~\ref{engss} plots the per-slot energy consumption of the proposed scheme when the landmarks are distributed in a
staircase (i.e., the two diagonally situated street blocks) or a square (i.e., the closed loop of a triangular street block),
with the spacing of $30$~m and $100$ m, respectively. Here, $T=600$ s, and $H=1000$~m or $1200$ m.
We see that the UAV consumes less energy when the landmark spacing or the UAV altitude is larger,
as the UAV embraces more flexibility to design an energy-efficient trajectory and meet the sensing requirements.
The per-slot energy consumption of the baseline approach is also plotted for the landmarks arranged on the square with the spacing
of $100$ m, and the UAV elevation is $H=1200$ m.
We see that the proposed scheme consumes dramatically lower energy than the baseline,
and prevents drastic fluctuations in the per-slot consumption of the UAV propulsion energy.

Fig.~\ref{figrndtwo} plots the UAV trajectory and per-slot energy consumption of the proposed and baseline schemes,
{\blue when the spacing between neighboring landmarks obeys a random uniform distribution within $[0, 100]$ m. The averaged results of 20 independent realizations of the landmark locations are plotted.}
While the UAV flies a longer distance under the proposed algorithm than it does under the baseline approach as shown in the top of the figure, the energy consumption is substantially lower and undergoes much smaller fluctuations under the proposed algorithm as shown in the bottom of the figure. 
This again corroborates the merits of the proposed scheme over the baseline scheme
in terms of energy efficiency and energy consumption fluctuation.

{\blue Last but not least, we examine the performance of the proposed BS-UAV bistatic SAR under different transmission bandwidths $B$
in the sub-6 GHz band and the mmWave band.
Table~\ref{tab.range} shows the range resolutions of the SAR, where $B$ is $50$~MHz, $100$~MHz, and $150$~MHz at the sub-6 GHz band, and $200$~MHz at the 
mmWave band. Here, $\theta_t$ is assumed to be $\pi/3$.
We see that the resolutions are always better than the expected shortest distance between any two objects, $d_{\min}$; 
in other words, the SAR can effectively distinguish ground objects.
Fig.~\ref{figturnbd} shows the UAV trajectory by the proposed and baseline schemes, where the landmarks are separated
by $200$ m, and the transmission bandwidth $B$ is $50$ MHz and $100$ MHz.
The other settings are consistent with those in Fig. 6.
By comparing Fig.~\ref{figturnbd} to Fig. 6, we see that the UAV trajectory has similar patterns while satisfying the sensing accuracy.
}

\section{Conclusion}\label{sec.con}
This paper proposed a novel framework of cellular-aided radar sensing with a BS-UAV bistatic SAR platform.
The trajectory of the UAV was optimized to minimize the propulsion energy while satisfying
the range and azimuth resolutions of sensing.
The trajectory planning problem was convexified and solved efficiently by utilizing the SCA and BCD methods.
Extensive simulations revealed that, in terms of energy efficiency and effective consumption fluctuation,
the proposed trajectory planning algorithm is superior to its alternative that minimizes
the flight distance of a cellular-aided sensing mission.
The energy saving can be as large as $55\%$ for the UAV, by running the proposed algorithm.
In the future, we will extend the BS-UAV bistatic SAR system to a fixed-wing UAV-based platform,
where a more sophisticated propulsion energy model and dynamic/mobility model will be taken into account.



\bibliographystyle{IEEEtran}
\bibliography{scuav_ref}

\end{document}